\documentclass[10pt,aps,prl,twocolumn,notitlepage,floatfix,shownopacs,preprintnumbers,nofootinbib,longbibliography]{revtex4-1}
\pdfoutput=1
\usepackage{amsmath,amsfonts,amsthm,bm,amssymb,mathtools,tabu}
\usepackage{graphicx}
\usepackage{gensymb}
\usepackage{mathbbol}
\usepackage{xfrac}
\usepackage[caption=false]{subfig}
\usepackage{epstopdf}
\usepackage{csquotes}
\usepackage[T1]{fontenc}
\usepackage{lmodern}
\usepackage{mathtools}
\usepackage[colorlinks=true,breaklinks=true]{hyperref}
\hypersetup{allcolors=[rgb]{0.0 0.0 0.6},linkcolor=[rgb]{0.8 0.0 0.4}}   
\usepackage{slashed}             
\usepackage{url}
\usepackage[detect-none]{siunitx}
\usepackage{color}  
\usepackage{comment}              
\usepackage[dvipsnames]{xcolor}

\DeclareTextCommand{\textprime}{\encodingdefault}{%
  \mbox{$\m@th'\kern-\scriptspace$}%
}
\DeclarePairedDelimiter\abs{\lvert}{\rvert}%
\DeclarePairedDelimiter\norm{\lVert}{\rVert}%
\makeatletter
\let\oldabs\abs
\def\abs{\@ifstar{\oldabs}{\oldabs*}}
\let\oldnorm\norm
\def\norm{\@ifstar{\oldnorm}{\oldnorm*}}
\makeatother

\allowdisplaybreaks

\bibpunct{[}{]}{,}{n}{}{,}

\providecommand{\abs}[1]{\lvert#1\rvert}
\providecommand{\norm}[1]{\lVert#1\rVert}

\providecommand{\href}[2]{#2}

\newcommand{\Real}{\mathbb{R}}

\renewcommand\({\left(}
\renewcommand\){\right)}
\renewcommand\[{\left[}

\newcommand{\GF}{G_{\rm F}}
\newcommand{\bp}{{\bf p}}

\newcommand{\br}{{\bf r}}
\newcommand{\bv}{{\bf v}}
\newcommand{\bk}{{\bf k}}
\newcommand{\bK}{{\bf K}}

\newcommand{\sH}{{\sf H}}
\newcommand{\sLambda}{{\sf\Lambda}}
\newcommand{\sM}{{\sf M}}
\newcommand{\sF}{{\sf F}}

\usepackage{mathtools}
\usepackage{xparse}
\makeatletter
\newcommand{\raisemath}[1]{\mathpalette{\raisem@th{#1}}}
\newcommand{\raisem@th}[3]{\raisebox{#1}{$#2#3$}}
\makeatother
\NewDocumentCommand{\dbar}{O{0pt} O{0pt}}{
  \ensuremath{\mathrlap{\raisemath{-0.48pt}{\hspace*{1.2pt}{\mathchar'26\mkern-9mu}}}d}%
}

\definecolor{blue}{rgb}{0.0,0.0,1.0}

\usepackage{tikz,xcolor,hyperref}

\definecolor{lime}{HTML}{A6CE39}
\DeclareRobustCommand{\orcidicon}{\hspace{-1mm}
	\begin{tikzpicture}
	\draw[lime, fill=lime] (0,0) 
	circle [radius=0.16] 
	node[white] {{\fontfamily{qag}\selectfont \tiny \,ID}};
	\draw[white, fill=white] (-0.0525,0.095) 
	circle [radius=0.007];
	\end{tikzpicture}
	\hspace{-3mm}
}

\foreach \x in {A, ..., Z}{\expandafter\xdef\csname orcid\x\endcsname{\noexpand\href{https://orcid.org/\csname orcidauthor\x\endcsname}
			{\noexpand\orcidicon}}
}

\begin{document}

\title{Collective Neutrino Flavor Instability Requires a Crossing}
	
\author{Basudeb~Dasgupta\orcidA{}}
\email{bdasgupta@theory.tifr.res.in}
\affiliation{Tata Institute of Fundamental Research,
             Homi Bhabha Road, Mumbai, 400005, India}

\date{November 5, 2021}

\begin{abstract} 
Neutrinos in supernovae, neutron stars, and in the early Universe may change flavor collectively and unstably, due to neutrino-neutrino forward-scattering. We prove that, for collective instability to occur, the difference of momentum distributions of two flavors must change sign, i.e., there is a zero \emph{crossing}. This \emph{necessary} criterion, which \emph{unifies} slow and fast instabilities, is valid for Hamiltonian flavor-evolution of ultra-relativistic Standard Model neutrino occupation matrices, including damping due to collisions in the relaxation approximation. It provides a simple but rigorous condition for collective flavor transformations that are believed to be important for stellar dynamics, nucleosynthesis, and neutrino phenomenology.

\end{abstract}

\preprint{TIFR/TH/21-15}

\maketitle
\emph{Introduction} -- Supernovae and neutron star mergers produce enormous numbers of neutrinos that carry away a bulk of the energy.
These neutrinos travel through the dense material of the star, and crucially influence stellar dynamics and nucleosynthesis~\cite{Colgate:1966ax, Bethe:1990mw, Janka:2016fox, Burrows:2020qrp, Thielemann:2017acv, Cowan:2019pkx}. The precise impact can depend on the flavor states of the neutrinos, because the different flavors interact with the background medium with unequal interaction rates. Naturally, a characterization of the neutrino flavor evolution in such environments is of interest and importance. 

The flavor evolution of dense neutrino clouds can be very complex~\cite{Duan:2010bg, Chakraborty:2016yeg, Tamborra:2020cul}. Neutrinos produced in the core of these sources initially remain trapped due to frequent collisions. They leak out via diffusion, before eventually free-streaming. If the density of background matter is high, flavor-mixing is suppressed due to forward and non-forward scatterings~\cite{Wolfenstein:1977ue, Mikheev:1987jp, Stodolsky:1986dx}. However, large collective flavor conversion can occur even for vanishingly small mixing. This novel \emph{instability} arises when neutrinos  forward-scatter off each-other and influence each-other's flavor evolution. As a result, the flavor evolution becomes intricately coupled, i.e., \emph{collective}, and creates nonlinear routes of exponentially growing flavor conversion.

In the last two decades, a lot of insight has been obtained into the origin and impact of a variety of collective flavor transformations.
Collective flavor transformations stem from neutrino-neutrino forward scatterings~\cite{Pantaleone:1992xh, Pantaleone:1992eq},  and come in three variants.
The simplest type are \emph{synchronized}, occurring for all neutrino energies with the average oscillation frequency $\langle \omega_E \rangle$, but not seeded by an instability~\cite{Kostelecky:1994dt}. These are usually suppressed in dense matter, though there is also the possibility of synchronized resonance that can cause large effects~\cite{Qian:1994wh}. The next type are \emph{slow instabilities}, leading to evolution with a frequency proportional to $\(\langle\omega_E\rangle\,\GF\,n_\nu\)^{\sfrac{1}{2}}$~\cite{Kostelecky:1993dm, Pastor:2001iu, Friedland:2003dv, Duan:2005cp, Duan:2006an,Hannestad:2006nj}, where $n_\nu$ is the neutrino density. At heart, these are analogous to the tipping of an inverted pendulum~\cite{Hannestad:2006nj, Raffelt:2007cb,Dasgupta:2009mg}.  Despite being dubbed slow, they are faster than the usual neutrino oscillations in vacuum/matter, or even the synchronized oscillations, because $n_\nu \gg \GF^{-1}\langle\omega_E\rangle$ deep in the star. As the neutrino density drops below $\GF^{-1}\langle\omega_E\rangle$, typically at a radius of a few $\times\,100$\,{\rm km} in a supernova, these slow instabilities tend to produce a swap of two flavors across broad range of energies~\cite{Raffelt:2007cb, Dasgupta:2009mg}. The edges of these swaps could appear as sharp \emph{spectral splits} in the energy spectrum and potentially observable in the signal reaching Earth. 
Finally, there are \emph{fast instabilities} that cause flavor evolution with a very high frequency proportional to $\GF n_\nu$~\cite{Sawyer:2005jk, Sawyer:2015dsa, Chakraborty:2016lct, Dasgupta:2016dbv,Izaguirre:2016gsx, Capozzi:2017gqd,Yi:2019hrp}. 
These can occur very deep in a star at radii of few $\times\,10$ km or so~\cite{Morinaga:2019wsv, Abbar:2019zoq, Glas:2019ijo}, and may impact stellar heating and nucleosynthesis in a more nontrivial fashion. Fast instabilities correspond to, at their simplest, motion in a quartic potential~\cite{Dasgupta:2017oko} or tipping of a pendulum~\cite{Johns:2019izj,Bhattacharyya:2020dhu}, and initially give wave-like propagation of flavor disturbance~\cite{Martin:2019gxb}. The eventual impact of these fast instabilities is not fully established yet, but a number of studies hint that they cause partial flavor equilibration in some range of neutrino velocities for all energies~\cite{Bhattacharyya:2020jpj, Johns:2020qsk,Wu:2021uvt,Richers:2021nbx}. This is called \emph{depolarization}, and may be a key observable of fast instability.  The large flavor conversion encoded in spectral splits or depolarization affects neutrino transport, thus affecting stellar evolution~\cite{Pejcha:2011en, Dasgupta:2011jf} and nucleosynthesis~\mbox{\cite{Duan:2010af, Xiong:2020ntn, Li:2021vqj}}, in addition to giving unique signals at detectors~\cite{Capozzi:2018rzl}.

A critical problem has been to determine the condition for collective instabilities. One belief has been that instabilities occur only if the flavor-difference distribution (FDD), i.e., the momentum distribution of the difference in the initial number densities of two neutrino-flavors, changes sign at some momentum~\cite{Dasgupta:2009mg, Dasgupta:2016dbv}. The importance of such FDD \emph{crossings} was first pointed out in a study of multiple spectral splits~\cite{Dasgupta:2009mg}. Several subsequent investigations have further strengthened this notion for slow~\cite{Banerjee:2011fj} and fast~\cite{Capozzi:2017gqd,Abbar:2017pkh,Capozzi:2019lso,Morinaga:2021zib} instabilities. Recently, a proof was proposed for necessary and sufficient condition for a \emph{fast} instability~\cite{Morinaga:2021vmc}. 

In this \emph{Letter} we show that ``FDD crossings are necessary for collective instability''. The argument is agnostic to whether the instability is slow or fast, to the number of neutrino flavors, and to whether damping due to collisions are present. This simple but rigorous criterion boxes-in the astrophysical circumstances where collective neutrino flavor transformations may be important. In the following, we derive a linearized evolution equation including damping to prove the above claim, and then conclude with a brief summary and remarks.

\emph{Dispersion Relation with Damping} -- We consider scenarios where neutrino flavor evolves as~\cite{Sigl:1992fn,McKellar:1992ja}
\begin{equation}\label{eq:EOM2a}
  v^\alpha\partial_\alpha\rho_\bp=-i\,[\sH_\bp,\rho_\bp]+{\sf C}_\bp\,,
\end{equation}
where $v^\alpha=(1,\bv)$ is the neutrino four-velocity with $\bv=\bp/|E|$, and a summation over the spacetime indices $\alpha=0,\ldots,3$ is implied. The $\rho_\bp$, ${\sH}_\bp$, and ${\sf C}_\bp$ are matrices in flavor-space, with $\rho_\bp$ (and $\bar\rho_\bp$) encoding the occupation density and flavor coherence for neutrinos (and antineutrinos) in their diagonal and off-diagonal entries; ${\sH}_\bp$ and ${\sf C}_\bp$ the Hamiltonian and collision matrices, respectively.  The problem is nonlinear because ${\sH}_\bp$ contains terms involving $\rho_\bp$ and $\bar\rho_\bp$, as does ${\sf C}_\bp$. The equation of motion (EoM) for $\bar\rho_\bp$ is the same except for a sign-change in the mass-mixing term in $\sH_\bp$.

In the following, we will derive a \emph{linearized} equation for the off-diagonal elements $\rho^{ij}_\bp$ in the flavor basis, where $i,j \in \{e,\,\mu,\,\tau\}$ in the usual three-flavor scenario. Our derivation remains essentially unchanged from ref.~\cite{Airen:2018nvp}, except for the inclusion of damping due to collisions.

The Hamiltonian matrix $\sH_\bp$ has the usual contributions
from neutrino mass-mixing as well as the refractive effects of other neutrinos and background leptons:
\begin{equation}\label{eq:EOM3b}
  \sH_{\bp}=\frac{\sM^2}{2E}+\sH^{\nu\nu}_\bp+\sH^{\rm bkg}_\bp\,.
\end{equation}
Explicitly, the neutrino-neutrino refractive term has the form
$\sH^{\nu\nu}_\bp=\sqrt{2}\GF\,v_\alpha\sF_\nu^\alpha$ with the
neutrino flux matrix $\sF_\nu^\alpha=\int \dbar\bp\,v^\alpha
  \(\rho_\bp-\bar\rho_\bp\)$, where $\dbar\bp={d^3\bp/(2\pi)^3}$. The ordinary matter contribution is $\sH^{\rm bkg}_\bp=\sqrt{2}\GF\,v_\alpha\sF_{\rm bkg}^\alpha$, which is diagonal and has the elements $(\sF_{\rm bkg}^\alpha)^{ii}=\int \dbar\bp\,u^\alpha_i\(f_{i,\bp}-\bar f_{i,\bp}\)$, for the $i^{\rm th}$ charged lepton with phase space distribution $f_{i,\bp}$ and a four-velocity $u_i^\alpha=(1,\bp/(\bp^2+m_i^2)^{1/2})$. With only at-rest electrons in the background, one finds the familiar matter potential ${\rm diag}\(2\sqrt{2}\GF n_e,\,0,\,0\)$ for three flavors. The expression used here is more general and includes other (anti)\,leptons as well as their currents. The mass-mixing term does not depend on $\bv$ and the refractive term does not depend on $E$, but only on $\bv$. We may define an overall matter effect caused by both neutrinos and charged leptons as
\begin{equation}
  \sH^{\rm matter,\,eff}=v_\alpha \sLambda^\alpha,
\end{equation}
where $\sLambda^\alpha={\rm diag}(\Lambda^\alpha_e, \Lambda^\alpha_\mu,\Lambda^\alpha_\tau)$ to represent the diagonal part of $\sqrt{2}\GF(\sF_{\rm bkg}+\sF_\nu)^\alpha$.

Collisions usually lead to damping of the off-diagonal flavor coherences~\mbox{\cite{Sen:2018put, Capozzi:2018clo,Shalgar:2020wcx,Martin:2021xyl, Johns:2021qby,Sasaki:2021zld}}. This is because off-diagonal elements can be enhanced only when neutrinos are produced or scattered into flavor non-diagonal states. Charged-current production/scattering leads to neutrinos in flavor eigenstates. Only neutral-current mediated scattering, with cross sections relatively suppressed by powers of $M_W^2/M_Z^2$, can lead to non-damping terms; E.g., a $Z$-mediated pair-production (scattering) can give a flavor-mixed neutrino in the final-state. We draw attention to the $d$, $c$, and $C$ terms in Sec.\,III of ref.~\cite{McKellar:1992ja}; see also refs.~\cite{Volpe:2013uxl, Vlasenko:2013fja, Kartavtsev:2015eva,Blaschke:2016xxt,Richers:2019grc}. Nearly isotropic distributions will suppress the non-damping effects of scattering. In general, however, collisions can produce coherence that may mimic an instability.

We will restrict this work to include collisional processes in the \emph{relaxation approximation}. For any pair of neutrino flavors, say $e$ and $\mu$, one has
\begin{equation}
{\sf C}^{e\mu}_{\bp}=-|\Delta^{e\mu}_{\bp}|\rho^{e\mu}_{\bp}\,,
\end{equation}
with the damping rate $|\Delta_{\bp}^{e\mu}|$ being non-negative. Similarly for any other pair of flavors.

In the limit of vanishing neutrino mixing, the linearized EoMs for the off-diagonal elements of $\rho_\bp$ (and their complex conjugates)
decouple, leading to equations of the form
\begin{eqnarray}\label{eq:EOM4}
  i\,v^\alpha\partial_\alpha \rho_{\bp}^{e\mu}&=&
  \(\frac{\sM^2_{ee}-\sM^2_{\mu\mu}}{2E}-i|\Delta^{e\mu}_{\bp}|
  +v_\alpha(\Lambda_e-\Lambda_\mu)^\alpha\)\rho_{\bp}^{e\mu}
  \nonumber\\[1ex]
  &&\hspace{-4em}-\sqrt{2}\GF\(f_{\nu_e,\bp}-f_{\nu_\mu,\bp}\)v^\alpha
  \int \dbar\bp'v_\alpha'\(\rho_{\bp'}^{e\mu}-\bar\rho_{\bp'}^{e\mu}\)\,,
\end{eqnarray}
and analogous for the other pairs of flavors.

In this approach the three-flavor system corresponds to three
independent two-flavor cases. There are three nontrivial cases only if the distributions of the three flavors are
different, as recently considered~\cite{Capozzi:2020kge}. Extension to more than three flavors is obvious.

All flavor coherence effects depend only on the difference of the original
neutrino distributions and the diagonal parts of all matrices in flavor space drop out. In particular,
we may write the effective two-flavor neutrino matrices of occupation numbers in the form
\begin{equation}\label{eq:s-define}
  \varrho_\bp^{e\mu}=\frac{f_{\nu_e,\bp}+f_{\nu_\mu,\bp}}{2}\,\mathbb{1}
  +\frac{f_{\nu_e,\bp}-f_{\nu_\mu,\bp}}{2}
  \begin{pmatrix}s_\bp&S_\bp\\S_\bp^*&-s_\bp\end{pmatrix}\,,
\end{equation}
whose off-diagonal element equals $\rho^{e\mu}_\bp$, where $s_\bp$ is a real number, $S_\bp$ a complex one, and $s_\bp^2+|S_\bp|^2=1$. To linear order in $|S_\bp|$, one has $s_\bp=1$, so in our linearized study we focus on the
space-time evolution of $S_\bp$ alone which holds all the information concerning
flavor coherence.

Defining the two-flavor matter effect through
$\Lambda^\alpha=(\Lambda_e-\Lambda_\mu)^\alpha$, the vacuum oscillation frequency
through $\omega_E=(\sM^2_{ee}-\sM^2_{\mu\mu})/(2E)$, and the damping as $|\Delta_\bp|$, the EoM in
equation~\eqref{eq:EOM4} becomes
\begin{eqnarray}\label{eq:EOM5}
  i\,v^\alpha\partial_\alpha S_\bp&=&
  \(\omega_E+v^\alpha \Lambda_\alpha-i|\Delta_{\bp}|\) S_{\bp}\nonumber\\[1ex]
 &&\hspace{2em} -v^\alpha \int \dbar\bp'v_\alpha'\(S_{\bp'}g_{\bp'}-\bar S_{\bp'}\bar g_{\bp'}\).
\end{eqnarray}
An analogous equation applies to the antineutrino flavor coherence
$\bar S_\bp$ with a sign-change of~$\omega_E$. Here we use the FDD, 
$g_\bp=\sqrt{2}\GF (f_{\nu_e,\bp}-f_{\nu_\mu,\bp})$ and
$\bar g_\bp=\sqrt{2}\GF (f_{\bar\nu_e,\bp}-f_{\bar\nu_\mu,\bp})$,
where we have absorbed $\sqrt{2}\GF$ for notational convenience.

These equations become more compact and
physically transparent in a convention where we
interpret antiparticles as particles with negative energy and describe
their FDD with negative occupation numbers. Thus the modes are labeled by
$-\infty<E<+\infty$ and their direction of motion $\bv$ with
$\bp=|E|\,\bv$. The two-flavor FDD is
\begin{equation}\label{eq:spectrum}
  g_\Gamma=\sqrt{2}\GF\begin{cases} f_{\nu_e,\bp}-f_{\nu_\mu,\bp}&\hbox{for $E>0$,}\\
  f_{\bar\nu_\mu,\bp}-f_{\bar\nu_e,\bp}&\hbox{for $E<0$,}
  \end{cases}
\end{equation}
with $\Gamma = \{E,\bv\}$. There is no sign-change in the definition of $S$. The EoM thus reads
\begin{equation}\label{eq:EOM6}
  \biggl(v^\alpha\(i\,\partial_\alpha-\Lambda_\alpha\)-\omega_E+i|\Delta_{\Gamma}| \biggr)S_{\Gamma}=
  -v^\alpha  \int d\Gamma'\, v_\alpha'\,g_{\Gamma'}\,S_{\Gamma'}\,,
\end{equation}
where the phase-space integration is over
\begin{equation}
  \int d\Gamma=\int_{-\infty}^{+\infty}\frac{E^2dE}{2\pi^2}\int \frac{d\bv}{4\pi}\,,
\end{equation}
with $\int d\bv$ an integral over the unit sphere, i.e., over the
polar and azimuthal angles of $\bp$.  

The vacuum oscillation frequency $\omega_E$, 
in this convention, automatically changes sign for antineutrinos.  For positive $E$, it is
positive for inverted mass ordering ($\sM_{ee}^2>\sM_{\mu\mu}^2$) and
negative for the normal mass ordering ($\sM_{ee}^2<\sM_{\mu\mu}^2$).

As usual, for a linear EoM we search for space-time dependent solutions of equation~\eqref{eq:EOM6} in terms of its independent Fourier components
\begin{equation}\label{eq:normalmode}
  S_{\Gamma,r}=\sum_K Q_{\Gamma,K}\,e^{-i(K_0 t-\bK\cdot\br)}\,,
\end{equation}
where $r^\mu=(t,\br)$ and $K^\mu=(K^0,\bK)$.
The quantity $Q_{\Gamma,K}$ is the eigenvector in $\Gamma$-space
for the eigenvalue $K$. To find the eigenmodes we insert the ansatz of
equation~\eqref{eq:normalmode} into equation~\eqref{eq:EOM6} and find
\begin{eqnarray}\label{eq:EOM7}
  \( v_\alpha k^\alpha-\omega_E +i|\Delta_{\Gamma}|\) Q_{\Gamma,k}=v_\alpha A^\alpha_k\,,
\end{eqnarray}
where $A^\alpha_k=-\int d\Gamma\, v^\alpha\,g_{\Gamma}\,Q_{\Gamma,k}$ and $k^\alpha=K^\alpha-\Lambda^\alpha$. Fully analogous to the fast-flavor case, we have shifted the original four-wavevector,  $K^\mu$, to the redefined four-wavevector, $k^\mu=(k^0,\bk)$, by subtracting the
matter-effect four-vector $\Lambda^\mu$. Solving the EoM in Fourier space allows the diagonal parts of all matter effect to be included as an origin-shift in the four-wavevector space.

In the absence of neutrino-neutrino interactions, the rhs of
equation~\eqref{eq:EOM7} vanishes and nontrivial solutions require
$v_\alpha k^\alpha-\omega_E+i|\Delta_{\Gamma}|=0$, i.e., the propagation relation ${\rm Re}\,k_0-\bv\cdot\bk=\omega_E$ and the damping ${\rm Im}\,k_0=-|\Delta_{\Gamma}|$, where each neutrino mode labelled by $\{E,\textbf{v}\}$ evolves independently.
In the presence of neutrino-neutrino
interactions, collective oscillations become possible where this
dispersion relation changes.
Therefore, we consider solutions
with $v_\alpha k^\alpha-\omega_E+i|\Delta_{\Gamma}|\not=0$ for any $\{E,\bv\}$
so that equation~\eqref{eq:EOM7} implies
\begin{equation}\label{eq:eigenvector}
  Q_{\Gamma,k}=\frac{v_\alpha A^\alpha_k}{v_\gamma k^\gamma-\omega_E+i|\Delta_{\Gamma}|}\,.
\end{equation}
Inserting this form on both sides of equation~\eqref{eq:EOM7} yields
\begin{equation}
  v_\alpha A^\alpha_k=-v^\alpha A^\beta_k\int d\Gamma'\,g_{\Gamma'}\,
  \frac{v'_\alpha v'_\beta}{v'_\gamma k^\gamma-\omega_{E'}+i|\Delta_{\Gamma'}|}\,.
\end{equation}
In more compact notation this can be written in the form
\begin{align}
 & v_\alpha\Pi_k^{\alpha\beta}A_{k,\beta}=0\,,\quad\hbox{where,}\,\label{eq:EOM8}\\  
&\Pi_k^{\alpha\beta}=h^{\alpha\beta}+\int d\Gamma\,g_{\Gamma}\,
  \frac{v^\alpha v^\beta}{v_\gamma k^\gamma-\omega_E+i|\Delta_\Gamma|}\,,\label{eq:EOM8b}
\end{align}
with $h^{\alpha\beta}={\rm diag}(+,-,-,-)$ being the metric tensor.
This equation must hold for any $v^\alpha$ and thus consists of
four independent equations $\Pi_k^{\alpha\beta}A_{k,\beta}=0$.
Nontrivial solutions require
\begin{equation}\label{eq:determinant}
{\cal D}(k) \equiv  {\rm det}\,\Pi^{\alpha\beta}_k=0\,,
\end{equation}
establishing a connection between the components of $k=(k^0,\bk)$, i.e.,
the dispersion relation of the system. It depends only on the neutrino
FDD $g_\Gamma$, which itself contains the neutrino density, the vacuum oscillation frequency
$\omega_E$, and the damping rate $|\Delta_\Gamma|$.

If the imaginary part of $k^0$ is positive, for any $k$ that satisfies equation (\ref{eq:determinant}), equation (\ref{eq:normalmode}) tells us that it leads to exponential growth of the off-diagonal flavor coherence between the two flavors under consideration, i.e., $S^{e\mu} \sim e^{t\,{\rm Im}\,k^0}$. In the limit of vanishing flavor-mixing, as relevant in dense matter, such flavor conversion is surprising and called a \emph{collective instability}.

\emph{Proof that Crossings are Necessary} -- Now we prove that collective instabilities can arise only if there is an FDD crossing. Technically, our proposition is that {if} any solution of the dispersion relation ${\cal D}(k)=0$ has ${\rm Im}\,k^0\equiv\sigma>0$ and  $\bk\in\Real^3$, {then} the {FDD} (i.e., $g_\Gamma$) cannot have the same sign everywhere. We will prove the proposition by contradiction, following \mbox{Morinaga}~\cite{Morinaga:2021vmc}, but cover a potential singular case.

In the following, we omit explicitly noting the $k$-dependence of the matrix $\Pi_k$ and its eigenvector $A_k$. Also we separate the real and imaginary parts of $k^0=\kappa+i\sigma$, where $\kappa,\sigma\in\Real$, and write the $\Pi$ matrix as
\begin{equation}
\Pi^{\alpha\beta}=M^{\alpha\beta}-iN^{\alpha\beta}\,,
\end{equation}
where $M$ and $N$ are real-symmetric matrices
\begin{align}
M^{\alpha\beta}&=h^{\alpha\beta}+\int d\Gamma\, g_\Gamma\, \frac{\(\kappa-\bv\cdot\bk-\omega_E\)v^\alpha v^\beta}{\(\kappa-\bv\cdot\bk-\omega_E\)^2+\(\sigma+|\Delta_{\Gamma}|\)^2}\,,\nonumber \\
N^{\alpha\beta}&=\int d\Gamma \,g_\Gamma\, \frac{\(\sigma+|\Delta_{\Gamma}|\)v^\alpha v^\beta}{\(\kappa-\bv\cdot\bk-\omega_E\)^2+\(\sigma+|\Delta_{\Gamma}|\)^2}\,.
\end{align} 
The matrix $N$ can be diagonalized by a real orthogonal matrix $O$ as
\begin{equation}
O^\alpha_\mu O^\beta_\nu N^{\mu\nu}=D^{\alpha\beta}\,,
\end{equation}
where $D$ is a diagonal matrix whose components are
\begin{equation}
D^{\alpha\alpha}=\int d\Gamma\,g_\Gamma\,\frac{\bigl(\sigma+|\Delta_{\Gamma}|\bigr)\,\bigl(O^\alpha_\mu v^\mu\bigr)^2}{\(\kappa-\bv\cdot\bk-\omega_E\)^2+\(\sigma+|\Delta_{\Gamma}|\)^2}\,.\label{eq:Dii}
\end{equation}
In this basis where $N$ becomes diagonal, the matrix $M$ becomes $\tilde{M}$ and 
the dispersion relation  ${\cal D}(k)=0$ becomes ${\rm det}\bigl(\tilde{M}-iD\bigr)=0$, which implies that there exists a nontrivial four-eigenvector $A$ such that
\begin{equation}
\tilde{M}^{\alpha\beta} A_\beta=+iD^{\alpha\beta} A_\beta\,.\label{eq:MDeqn}
\end{equation}
Note that $\tilde{M}-iD$ is a complex-symmetric matrix, so in general $A$ is a complex vector. We multiply the above equation by ${A}^*_\alpha$ and sum over $\alpha$ to get
\begin{equation}
\tilde{M}^{\alpha\beta}{A}^*_\alpha A_\beta=+iD^{\alpha\beta}{A}^*_\alpha A_\beta\,,\label{eq:Avec}
\end{equation}
whose complex conjugate is given by
\begin{equation}
\tilde{M}^{\alpha\beta}A_\alpha A^*_\beta=-i D^{\alpha\beta}A_\alpha A^*_\beta\,.
\end{equation}
Using  the fact that $\alpha$ and $\beta$ are dummy indices and can be renamed  $\beta$ and $\alpha$, respectively, and that $\tilde{M}$ is symmetric, i.e., $\tilde{M}^{\beta\alpha}=\tilde{M}^{\alpha\beta}$, we get
\begin{equation}
\tilde{M}^{\alpha\beta}A^*_\alpha A_\beta=-i D^{\alpha\beta} A^*_\alpha A_\beta\,.\label{eq:Avec2}
\end{equation}
Subtracting equation (\ref{eq:Avec2}) from equation (\ref{eq:Avec}) gives
\begin{equation}
\sum_\alpha D^{\alpha\alpha}|A_\alpha|^2=0\,.\label{eq:final}
\end{equation}
In the equation (\ref{eq:final}) above, $|A_\alpha|^2$ are non-negative and not all of them vanish. As proposed, we have $\sigma>0$ and $g_\Gamma$ has the same sign everywhere, so equation (\ref{eq:Dii}) dictates that all $D^{\alpha\alpha}$ have the same signature as $g_\Gamma$.

There would appear to be two possibilities for \mbox{equation}~(\ref{eq:final}). First, the singular case where $D^{\alpha\alpha}=0$ for \emph{all} $\alpha$ for which $|A_\alpha|^2\neq0$. However, in this case equation (\ref{eq:Dii}) requires that  for those $\alpha$, the integral of $\bigl(O^\alpha_\mu v^\mu\bigr)^2$ times an \emph{everywhere-same-sign} function vanishes identically. This is possible only if $\bigl(O^\alpha_\mu v^\mu\bigr)^2=0$ for all points in $\Gamma$ or if $g_\Gamma=0$. That is, the same $O$ makes the $\alpha$-component of any $v$ vanish or that there are no collective effects at all, respectively. These are either impossible or trivial, and therefore excluded.
Second is the non-singular case, where $D^{\alpha\alpha}\neq0$ for \emph{some} $A_\alpha\neq0$. In this case, in equation (\ref{eq:final}) at least one term is nonzero and all terms are of same sign. But then equation~(\ref{eq:final}), which algebraically followed from our original assumptions, cannot be satisfied! The only resolution is that $g_\Gamma$ must change sign if there exists a $\sigma>0$. This completes the proof of the proposition. As a corollary, setting $\omega_E \to 0$ and $\Delta_\Gamma\to0$, one recovers the necessary condition for collisionless fast instability~\cite{Morinaga:2021vmc}.\\[-3ex]

\emph{Discussion {\&} Outlook} -- We have shown that collective flavor instability \emph{requires} an FDD crossing in momentum space. Observation of signatures of collective neutrino flavor instabilities, such as spectral splits, depolarization, and their impact on stellar heating and nucleosynthesis, will therefore provide information on neutrino distributions deep inside stars. The criterion presented in this work gives a rigorous foundation for this physical expectation. It also makes it eminently
sensible to rule out collective flavor instabilities in supernova
simulations by simply ruling out FDD crossings~\cite{Dasgupta:2018ulw, Abbar:2020fcl, Johns:2021taz}. We note that this criterion \emph{unifies} the origin of slow and fast instabilities, and is distinct from the crossing criterion for fast instabilities where one demands a crossing in the energy-integrated FDD~\cite{Morinaga:2021vmc}. Also, note that the proof does not depend on any detailed features of $g_\Gamma$; thus it cannot be violated by invoking azimuthally non-symmetric distributions, for example.  

We now outline some limitations of the necessary criterion. Our framework considers evolution of the occupation matrices, \mbox{$\rho\sim\langle a^\dagger a\rangle$} and \mbox{$\bar\rho\sim\langle b^\dagger b\rangle$}, related to field-theoretic expectations of the fermion bilinears. Other correlations, e.g., $\langle a^\dagger b\rangle$, encode coupling of helicity, spin, lepton-number. While much tinier, these demand an expanded Hilbert space~\mbox{\cite{Volpe:2013uxl, Vlasenko:2013fja, Kartavtsev:2015eva,Blaschke:2016xxt,Richers:2019grc}}, which could invalidate the proof. Similarly, going beyond the mean-field approximation~\mbox{\cite{Patwardhan:2019zta, Roggero:2021asb}} needs a more sophisticated treatment. Beyond the Standard Model, even certain kinds of forward interactions render the linearized EoM to not remain an eigenvalue equation~\cite{Dighe:2017sur}, and the proof ceases to apply. Also, note that our analysis is limited to \emph{linear} instability. Finally, within the assumed framework, the collision term could have non-damping parts that we have ignored.  These extensions may reveal novel collective oscillation effects that are not accounted by the crossing criterion that was proved here.

One could ask if an FDD crossing is also \mbox{\emph{sufficient}} for a collective instability? In the fast limit, it has been proposed that a crossing in the energy-integrated FDD spectrum, must lead to some solution $k$ of the dispersion relation ${\cal D}(k)=0$ with a complex $k^0$ and real $\bk$~\cite{Morinaga:2021vmc}. The proof in effect recasts the dispersion relation as a quartic polynomial, which we couldn't reproduce. Note also, even if a crossing guarantees instability, the rate of the instability can be very small.

The exploration of collective effects over the past three decades has revealed many novel phenomena. The proposed necessary criterion hopefully provides an organizing principle, and its violations may reveal yet newer secrets of collective neutrino flavor transformations.

\emph{Acknowledgements} -- I thank S. Abbar for organizing the ``Collective Oscillation Exchange'' online journal club, where T. Morinaga presented a seminar on his proof~\cite{Morinaga:2021vmc}. That a generalization might be possible was suggested by the present author, in the Q{\emph \&}A following the seminar.  It is my pleasure to thank S.\,Abbar, T.\,Morinaga, G.\,Raffelt, and R.\,Sawyer for detailed discussions. I also thank A.\,Dighe, L.\,Johns, and M.\,Sen for helpful exchanges. This work is supported by the Dept.~of Atomic Energy~(Govt.~of~India) research project RTI 4002, the Dept.~of Science and Technology~(Govt.~of~India) through a Swarnajayanti {Fellowship}, and by the Max-Planck-Gesellschaft through a {Max Planck Partner Group}.

\linespread{0.95}

\providecommand{\href}[2]{#2}

\begingroup

\endgroup

\end{document}